# The Physics of Magnetars II - The Electron Fermi Energy of and the Origin of High X-ray Luminosity of Magnetars

Qiu-he Peng[1,2,*], Jie Zhang[3], Chih-kang Chou[2], Zhi-fu Gao[4]

[1]Department of Astronomy, Nanjing University, China
[2]National Astronomical Observatory, Chinese Academy of Sciences, China
[3]Department of Physics, West-China Normal University, China
[4]Xinjiang Astronomical Observatory, Chinese Academy of Sciences, China

**Abstract**   In this paper we discuss in detail the quantization of Landau energy levels of a strongly magnetized and completely degenerate relativistic electron gas in neutron stars. In particular, we focus on the Fermi energy dependence of the magnetic field for a relativistic electron gas in the superstrong magnetic field of magnetars. We would like to point out that some of the results concerning the microscopic number density of states of a strongly magnetized electron gas given by well-known statistical physics text books are incorrect. The repeated use of these results in the last five decades probably seriously affects the physics of neutron stars and magnetars. The quantization of Landau energy levels is accurately delineated in terms of the Dirac -δ function. Relatively simple calculation shows that the Fermi energy of a relativistic electron gas in magnetars with superstrong magnetic fields ($B > B_{cr}$, here $B_{cr}$ is the Landau critical magnetic field, $B_{cr} = 4.414 \times 10^{13} gauss$) increases with magnetic field strength as $B^{1/4}$. On the basis of this simple but important new result we are leading naturally to propose a new mechanism for the production of high X-ray luminosity from magnetars.

## 1. Introduction

The motion of a free electron along an external magnetic field is not affected by the field whereas the motion of the electron across the magnetic field is circular due to the action of the Lorentz force with gyration frequency $\omega_c$,

$$\omega_c = eB/m_e c, \qquad (1)$$

Where $m_e$ is the rest mass of the electron, c is the speed of light, e is the charge of the electron. We have

$$\hbar \omega_c = 2\mu_B B, \qquad (2)$$

where $\mu_B$ is electron Bohr magnetic moment, $\mu_B = e\hbar/2m_e c = 0.927 \times 10^{-20} ergs/gauss$.

The gyration radius of the electron, $r$, is related to the momentum of the free electron perpendicular to the magnetic field, $p_\perp$, as $r = cp_\perp/eB$. Thus, the gyration radius is smaller for stronger magnetic fields. In a magnetized electron gas at equilibrium temperature, $T$, The microscopic quantum mechanism effect must be considered when the gyration radius, r is smaller than the Debye radius, $r_D = kT/\sqrt{4\pi n_e e^2}$, where $n_e$ denotes the electron number density. For a magnetized electron gas with relatively low density and moderate magnetic field strength the microscopic quantum mechanical behavior of the system may be delineated in terms of the non-relativistic Landau orbital quantization. The wave function of the Schrödinger equation for the electrons $\psi(\rho,\varphi,z) = \chi(\rho)e^{im\varphi}e^{ikz}$ with magnetic quantum numbers $m = 0, \pm 1, \pm 2,...$ , $k = p_z/\hbar$, $-\infty < k < \infty$, $p_z$ is the electron momentum along the external magnetic field.

The corresponding Schrödinger equation for the radial wave function, $\chi(\rho)$ may be written as

$$\{-\frac{\hbar^2}{2m_e}(\frac{\partial^2}{\partial \rho^2} + \frac{1}{\rho}\frac{\partial}{\partial \rho} - \frac{m^2}{\rho^2}) + \frac{1}{2}m_e \omega_L^2 \rho^2\} \chi(\rho) = E' \chi(\rho),$$

(3)

Where

$$E' = E - m\hbar\omega_L - \hbar^2 k^2/2m_e \quad , \quad (4)$$

$$\omega_L = \omega_c/2 = eB/2m_e c \quad . \quad (5)$$

Here $E$ is the energy proper value of the electron. For a fixed $p_z$, the Schrödinger equation may be solved by using the boundary condition that the radial wave function must be finite at infinity and the corresponding series expansion for the radial wave function must terminate. The procedure mentioned above then naturally leads to the energy levels of the Landau orbital quantization. We would like to emphasize that the continuous momentum $p_z$ along the magnetic field is our first choice for the good quantum numbers. For a fixed $p_z$ we then determine the other orbital quantum numbers $n = 0,1,2,3,...$ by solving the Schrödinger equation as just stated above. We now summarize the important results as follows.

The momentum, $p_z$ of the free electron along the magnetic field can change continuously whereas the momentum of the free electron across the magnetic field must be quantized. The energy for a free electron in a strong and uniform magnetic field may be cast in the form[1]

$$E = \frac{p_z^2}{2m_e} + (n + \frac{1}{2} + \sigma)\hbar\omega_L \quad (6)$$

The density of states of the electron gas in the momentum interval $p_z \to p_z + dp_z$ is given by[1]

$$\frac{eB}{4\pi\hbar^2}\frac{dp_z}{c}. \quad (7)$$

The equation (7) is obtained on the basis of non-relativistic theory for a gyration electron in a magnetic field.

However for applications to neutron stars and magnetars, we must consider more extreme cases of very high density and ultra-strong magnetic field. In particular, it is well known that fully relativistic theory must be used to depict the gyrating electrons in strong magnetic field that approach or exceed the critical field strength $B_{cr}$ determined by $2\mu_e B_{cr}/m_e c^2 = 1$ or

$B_{cr} = m_e c^2/2\mu_e = m_e^2 c^3/e\hbar \approx 4.414 \times 10^{13} Gauss$ .

We note that the energy of the magnetic moment of the electron in external magnetic fields would exceed the rest mass energy of the electron if the field strength of the external magnetic field exceeds the critical field strength $B_{cr}$. Moreover, at such ultra strong magnetic fields the gyrating energy $\hbar\omega_L$ also exceeds the rest mass energy of the electron. It is clear that the non-relativistic theory of Landau orbital quantization must be modified. In addition, the electron gas in neutron stars and white dwarfs is almost completely degenerate and the corresponding Fermi energy in such stellar objects is much larger than the electron rest mass energy, $E_F(e) >> m_e c^2$. It is well known that for such super-strong magnetic fields the appropriated relativistic Dirac electron theory yields the correct Landau energy level,

$$(\frac{E}{m_e c^2})^2(p_z, B, n, \sigma) = 1 + (\frac{p_z}{m_e c})^2 + (\frac{p_\perp}{m_e c})^2$$
$$= 1 + (\frac{p_z}{m_e c})^2 + (2n+1+\sigma)\frac{2\mu_e B}{m_e c^2},$$
$$= 1 + (\frac{p_z}{m_e c})^2 + (2n+1+\sigma)b$$

(8)

Where $b = B/B_{cr}$, i.e. $b$ is a parameter denoting the ratio of the magnetic field with the critical field strength.

In Section II, we use the relativistic Dirac electron theory to derive the Fermi energy for the electron gas in the presence of a superstrong magnetic field. The resulting expression for the magnetic field dependence of the Fermi energy will be used in Section III to propose a mechanism for the production of high X-ray luminosity of magnetars. Finally, in Section IV we discuss other important applications and the physics significance of main result.

## 2. The Fermi Energy of the Electron Gas in Super-strong Magnetic Fields of Magnetars

The Fermi sphere in the weak magnetic field will be shaped into a Landau column in the strong magnetic field. In the x-y plane perpendicular to magnetic field, electrons are populated in discrete Landau levels, although $p_z$ changes continuously. When the magnetic field is weak the effect of the Landau orbital quantization is negligible, and the Fermi sphere for the highly degenerate electron gas is isotropic in the momentum space basically.

In case of high temperature and/or low density satisfied the following condition $Y_e \rho(g \cdot cm^{-3})/10^4 \leq 2.4(T/10^8 K)^{3/2}$ ($Y_e$ is the electron fraction) for the classical Boltzmann gas, we have : a) the quantum numbers of Landau level n, may approach infinity theoretically; b) the probability that an electron occupies at a high energy level in a strong magnetic field will be very small because the probability is proportional to the Boltzmann factor $\exp\{-(E/kT)\}(E >> kT)$, and it can be really ignored. For simplification, one can assume that the free electrons whose motion is perpendicular to the magnetic field are almost completely concentrated to a few states with lowest energy levels (n = 0,1,2,3).

However, the behavior of the Fermi degenerate electron gas is different with that of the classical Boltzmann gas in the strong magnetic fields. When the magnetic field is very strong, the quantization effect of the Landau levels along the direction perpendicular to magnetic field is very strong. The Fermi sphere deforms into a series of quantization Landau cylinder(see Fig.1). For a completely degenerate Fermi electron gas, every micro-state inside of the Fermi sphere is occupied by one electron due to the Pauli exclusion principle. And no any electron is out of the Fermi surface . Giving the momentum $p_z$, of the electron along the magnetic field, the electron momentum perpendicular to the magnetic field $p_\perp$, is the radius of the cross section for the series of Landau cylinder with quantum number $n$ (see the Fig. 1), where $n = 0,1,2,...n_{max}$. The maximum of the quantum numbers is

$$n_{max}(p_z, b, \sigma = -1) = Int\{\frac{1}{2b}[(\frac{E_F}{m_e c^2})^2 - 1 - (\frac{p_z}{m_e c})]\}$$

$$n_{max}(p_z, b, \sigma = +1) = Int\{\frac{1}{2b}[(\frac{E_F}{m_e c^2})^2 - 1 - (\frac{p_z}{m_e c})] - 1\}$$

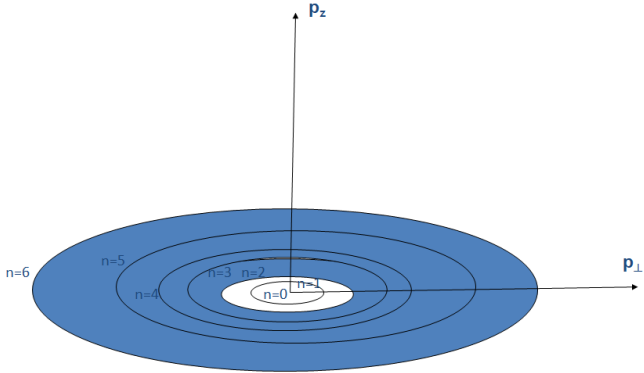

**Figure 1.** Landau energy quantization

In this article we make the following approximate expression reasonably.

$$n_{max}(p_z, b) \approx n_{max}(p_z, b, \sigma = -1) \approx n_{max}(p_z, b, \sigma = +1)$$
$$\approx \frac{1}{2b}[(\frac{E_F}{m_e c^2})^2 - 1 - (\frac{p_z}{m_e c})]$$
(9)

The maximum number, $n_{max}(p_z, b)$, is smaller for a larger magnetic field.

$n_{max}(p_z, b)$ is a function of the magnetic field and it evidently decreases with the magnetic field increasing.

Giving number density of electrons in the interior of neutron stars, the stronger the magnetic field is, the fewer the number of states occupied by electrons across the magnetic field. Therefore, the maximum momentum along the magnetic field $p_z(max)$ will increase with increasing magnetic field in the case that the total number of electrons in a unit volume is fixed. Thus the electron Fermi energy, $E_F(e)$, will increase with increasing magnetic field also where the relation $E_F(e) \approx cp_F = cp_z(max)$ for the ultra-relativistic electrons. Thus we come a very important conclusion naturally: **the Fermi energy of the electron gas increases with increasing magnetic field in the magnetars.**

## 3. A Misleading Approach regarding the Fermi Energy of the Electron Gas in Super-strong Magnetic Field

Unfortunately, we are surprised to find that the generally accepted idea is just on the contrary, namely, " the stronger the magnetic field, the lower the electron Fermi energy''. This popular idea has propagated in almost all the relevant literature (e.g. refs.[2-5]) for more than four decades. By tracking this big discrepancy we find that the authors of these papers directly or indirectly cite the internationally popular classic statistical physics textbooks (see e. g., Refs [6] and [7]) , in which the authors give a formula of calculating the microstate number for the electron gas in the interval of the electron momentum along the direction of the magnetic field, $p_z \to p_z + dp_z$.

In this section, we shall argue in detail that the relative formula given in Refs [6] and [7] is not valid in the case with super-strong magnetic field of magnetars.

In the case of the non-relativistic electron gas with the weak magnetic field, the microstate density of electron gas in the momentum interval, $p_z \to p_z + dp_z$, is given by Eq.(7)[1] mentioned in §I , which is obtained by solving the equation of non-relativistic Lamor gyration motion. However, the Eq.(7) is no longer suitable for the case with super-strong magnetic field, $B > B_{cr}$ due to the energy of gyration motion of the electron being ultra relativistic, $\hbar\omega_L > m_e c^2$. In case of super-strong magnetic field the Eq.(7) should be recalculated. However, it is very difficult to seek the solution of the relativistic gyration motion equation. The way taken by Refs. [6] and [7] is one of methods recalculating the microstate number of the electron gas. However there is a serious mistake in their calculation. They use the following method to calculate the statistical weight (degeneracy) of the Landau level $n$. The possible microscopic state number density in the interval of momentum $p_z \to p_z + dp_z$ along the magnetic field direction for the electron gas is given by

$$N_{phase}(p_z) = h^{-2}\int dp_x dp_y = h^{-2}\pi(p_\perp^2)_n^{n+1} = 4\pi m\mu_B B/h^2$$
(10)

In their calculation. The result of the microscopic state number for the level (n +1) in the plane perpendicular to magnetic field is essentially equal to the area of the torus between $p_\perp(n) \to p_\perp(n+1)$ (See Fig.2). However, this is obviously inconsistent with the key concept of Landau orbital quantization under a strong magnetic field. Really no quantum state between $p_\perp(n) \to p_\perp(n+1)$ in the momentum space is allowed according to the idea of Landau orbital quantization . Therefore, their result is wrong in case of super-strong magnetic field.

Following is how they calculate the number of microscopic states: the number density of microscopic states is

$$N_{phase} = \int_0^{p_F} N_{phase}(p_z) dp_z = \frac{eB}{4\pi\hbar^2} \frac{E_F}{c^2}, \quad (11)$$

by using Eq.(10). According to the Pauli Exclusion Principle, the number density of microscopic states is equal to electrons number density in a complete degenerate electron gas, $N_{phase} = n_e = N_A \rho Y_e$, where $Y_e$ and $\rho$ are electron fraction (5-8%) and mass density. This implies: $E_F(e) \propto B^{-1}$.

This leads to popular idea that the electron Fermi energy decreases with increasing magnetic field.

Because this result is derived from the formula Eq. (10), which is wrong in case of super-strong magnetic field, we argue that the popular idea, the electron Fermi energy decreases with increasing magnetic field, is also wrong. The correct idea is suggested in the paragraph§II.

Unfortunately, this incorrect result has been used by many authors ([2]-[5]) for the case of the strong magnetic field in neutron stars during the last four decades. For example, In the Page 12 of Ref. [2], authors clearly cited the result of the statistical physics textbooks edited in 1965. Many relevant papers directly or indirectly cited reference [3],[4] as a starting point since 1991.

There is another questionable approach in the reference [6] and [7]. They do not use the Dirac's δ-function in calculating the integral of electron momentum perpendicular to the magnetic field. The Landau level quantization is introduced by hand in the end of the calculation. It is mathematically no naturally.

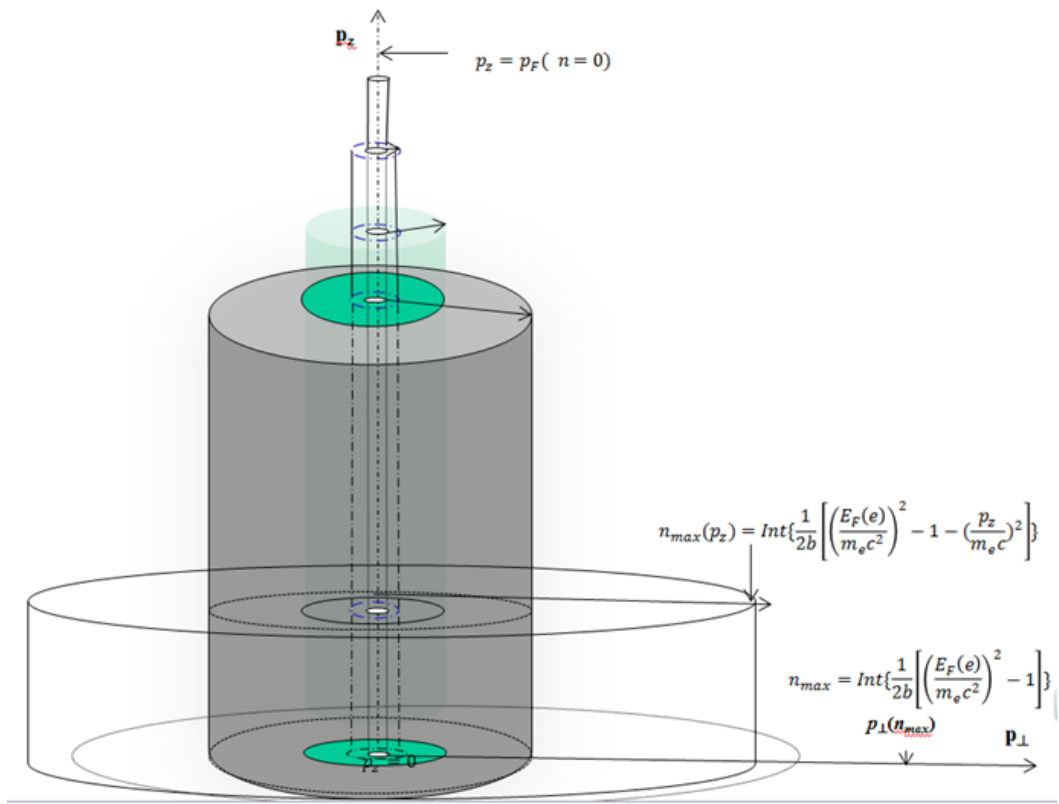

**Figure 2.** Electron population in strong magnetic field

## 4. A Questionable Approach Regarding the Calculation of Physical Behavior of Electron Gas in the Neutron Stars with the Strong Magnetic Field

Some calculating method for the behavior of the electron gas in the neutron stars with the strong magnetic field is also questionable in the relevant literature (e.g. refs.[2-5]). In this section, we shall show that their mathematical calculation methods are inappropriate.

First, the authors of these articles (e.g., [3], [4]) spread an infinite series through the Landau level quantum numbers ($n_L = 0,1,2,3...,\infty$) to calculate some physical quantities, e.g. both the number density of the electron gas and the electron pressure, $n_e, P_e,...$ in the neutron star with the strong magnetic field.

Second, for a given Landau level quantum number($n_L$), the authors calculate the infinite integral of the electron's momentum ($p_z$) along the direction of the magnetic field, e.g., Eq. (2.5) in the literature [3] and Eq. (6.1) and Eq. (6.3) in the literature [4]. That means, the calculating steps in these literature(e.g., [2]-[5]) are as following: given the Landau level quantum number($n_L$), firstly, the authors calculate an infinite integral for the continuous momentum $p_z$ of the electron along the magnetic field, then secondly they make the summation of the infinite series by Landau level quantum number ($n_L = 0,1,2,3...,\infty$). But this calculation order is just opposite to that of the Landau original idea.

As we have emphasized in §I of this paper that the continuous momentum $p_z$ of the electron along the magnetic field $p_z$ is preferred good quantum numbers. For a fixed $p_z$, the series of Landau orbital quantum numbers $n = 0,1,2,3,...$ are then determined by solving the Schrödinger equation. Thus we should follow the Landau calculating order: the first operation is summation of the infinite series for the Landau level quantum number ($n_L = 0,1,2,3...,\infty$), then taking the result of the summation as an integrand, we then calculate the infinite integral for the continuous momentum $p_z$ of the electron along the magnetic field. However, the calculation steps adopted by the authors of the relevant literature(e.g., [2]-[5]) are just inverse of the steps for the Landau idea.

The key question is that can the order of these two operations be exchanged? Our answer is no! The reason is very simple. From the well-known calculus, we know, the exchange of the order for the summation of the infinite series and for the calculation of the infinite integral has to satisfy very strict conditions such as "absolute convergence of infinite series" and the condition of both absolute convergence and uniform convergence for the infinite integral of the continuous momentum $p_z$ of the electron along the magnetic field. However no any discussion of these prerequisite conditions has been found in the relevant literature(e.g., [2]-[5]). In these papers, the authors exchange the calculation order directly and they use certain physical constraints to limit the infinite integral to a limit integral. This approach leads to very complex and tedious calculations. In fact, all the relevant theories so far have not been tested by the observations yet, and it is especially hard to explain the observed phenomena of magnetars or neutron stars in strong magnetic field by their theories. Although their calculation steps are questionable as we mentioned above, the key error of these articles discussing the effect of the strong magnetic field is that these authors have cited the Eq.(7) above for the calculation of the number of microscopic states of the electron gas in the popular statistical physics textbooks[6] and [7]).

## 5. Magnetic Field Dependence of the Fermi Energy for Relativistic Electrons: Our Method and Results

When the magnetic field is weak, $b << 1$, we have $n_{max}(p_z,b) >> 1$ according to the Eq.(9). That means that as long as the magnetic field is obviously weak, $n_{max}$ is very large. It returns to the non-relativistic Landau case. However, for the completely degenerate and moderate relativistic electron gas in the normal neutron stars, the calculation of the total number of microstates is very complex (attributed to seek moderately relativistic Larmor cyclotron motion equation of the solution has not been solved up to date). We don't discuss this situation. We are only discussing the case with superstrong magnetic field, $\mathbf{B} > B_{cr}$ (i.e. $b > 1$).

As well known, the microstate number in the six dimensional phase space volume element, $d^3x d^3p$ is $\delta N_{phase} = \frac{1}{h^3} dxdydzdp_x dp_y dp_z$. To re-examine the method of calculating the microscopic state number density for the electrons in strong magnetic fields, we introduce the Dirac-δ function to rigorously describe the Landau orbital quantization along the direction perpendicular to the magnetic field in the momentum space. The total number density for the electrons in a strong magnetic field is given by

$$N_{phase} = 2\pi (\frac{m_e c}{h})^3 g_0 \int_0^{p_F/m_e c} d(\frac{p_z}{m_e c}) \{ \sum_{\sigma=\pm\frac{1}{2}} \sum_{n=0}^{n_{max}(p_z,b,\sigma)} g_n \times$$

$$\int_0^{p_F/m_e c} \delta[\frac{p_\perp}{m_e c} - \sqrt{2(n+\frac{1}{2}+\sigma)b}](\frac{p_\perp}{m_e c}) d(\frac{p_\perp}{m_e c}) \}$$

(12)

Where $g_n$ is the degeneracy of electron spin. $g_0 = 1$ for $n=0$, and $g_n=2$ for $n \geq 1$.

By use of the behavior of Dirac δ - function the integral may be easily rewritten as

$$N_{phase} = 2\pi(\frac{m_e c}{h})^3 g_0 \int_0^{p_F/m_e c} d(\frac{p_z}{m_e c})\{\sum_{n=0}^{n_{ma}(p_z,b,\sigma=-1)} \sqrt{2nb}$$

$$+ \sum_{n=1}^{n_{ma}(p_z,b,\sigma=+1)} \sqrt{2(n+1)b}\}$$

We put the approximate Eq.(9) into the up limit of math series summation in the integrand. due to $n_{\max}(p_z,b) \gg 1$, the math series summation for the Landau level discrete quantum number n, may be approximately substituted by a continuous variable integral in general. Thus we have

$$N_{phase} = \frac{4\pi}{3b}(\frac{m_e c}{h})^3 g_0 \int_0^{p_F/m_e c} [(\frac{E_F}{m_e c^2})^2 -$$

$$-1-(\frac{p_z}{m_e c})^2]^{3/2} d(\frac{p_z}{m_e c}) \quad (13)$$

$$= \frac{4\pi}{3b} g_0 I (\frac{m_e c}{h})^3 (\frac{E_F}{m_e c^2})^4$$

Where $I = \int_0^1 (1-t^2)^{3/2} dt = 3\pi/16$, $h/m_e c = \lambda_e$ is the Compton wavelength of electrons. The number of microscopic states in the unit volume is equal to the number density of completely degenerate electrons by the Pauli Exclusion Principle, i.e., $N_{phase} = n_e = N_A \rho Y_e$. Consequently we obtain

$$E_F(e) \approx 42.9 (\frac{Y_e}{0.05})^{1/4} (\frac{\rho}{\rho_{nuc}})^{1/4} (\frac{B}{B_{cr}})^{1/4} \text{ MeV}. \quad (14)$$

This means that the Fermi energy of the relativistic electrons in strong magnetic fields is proportional to the one forth power of the field strength.

## 6. A Physical Mechanism of High x-ray Luminosity for Magnetars

In terms of our new result for the Fermi energy with strong magnetic field, Eq.(14), we may easily explain the physical origin for high x-ray luminosity of magnetars. The main idea is as follows: since magnetic fields in magnetars are much stronger than the critical magnetic field, $B_{cr}$, the Fermi energy of the electrons in magnetars is very high. Owing to the Eq.(14), when the Fermi energy of the relativistic electrons significantly exceeds that of the (non-relativistic) neutrons ( $E_F(n) \approx 60 MeV$ ), some electrons can be captured by the protons near the Fermi surface of the protons. In this case, a reaction of the electron capture by a proton, $e^- + p \rightarrow n + \nu_e$, will happen, although the reaction is forbidden in the normal neutron stars, due to the energy of the outgoing neutron being lower than the Fermi energy of the neutrons.

Some neutrons will be emitted and the resulting energy of the outgoing neutrons ( > 60 MeV) is much higher than the binding energy( $\approx 0.045 MeV$ ) of the $^3P_2$ neutron Cooper pair. The outgoing neutrons can destroy the $^3P_2$ neutron Cooper pairs by strong nuclear interactions: ($n + (n\uparrow, n\uparrow) \rightarrow n + n + n$). We note that the binding energy of the $^3P_2$ neutron Cooper pair is compensated by the energy of the outgoing neutron.

When the Cooper pairs are destroyed, the spins of the two neutrons forming the $^3P_2$ neutron Cooper pairs are no longer parallel. Thus they are in the state of random thermal motion and in a chaotic state. In this case, the magnetic moments of the $^3P_2$ neutron Cooper pairs with twice of the neutron anomalous magnetic moment, $2\mu_n$, would disappear. And then the induced magnetic moment generated by the $^3P_2$ Cooper pair also disappear. The magnetic moments of the $^3P_2$ neutron Cooper pairs have the tendency in the reverse direction of the magnetic field. The energy of the magnetic moments, $2\mu_n B$, could be transformed into the chaotic thermal energy when the destruction of the Copper pairs take place. The heat released by each $^3P_2$ neutron Cooper pair is $kT = 2\mu_n B \approx 10 B_{15}$ keV, where $B_{15} = B/10^{15}$ gauss. The thermal energy then convert into the x-ray radiation of the magnetars. When the entire $^3P_2$ neutron Cooper pairs are destroyed, the total energy released will be

$$E = qN_A m(^2P_3) \times 2\mu_n B \approx 1 \times 10^{47} \, m(^2P_3)/(0.1 m_{Sun}) ergs, \quad (15)$$

where $m(^3P_2)$ is the total mass of the superfluid region with anisotropic $^3P_2$ neutron pair, $N_A$ is the A'vgadro content, and $q$ is the ratio for the number of neutrons combined into $^3P_2$ neutron Cooper pairs to the total number of neutron in the superfluid region (about 8.7%)[8]. According to the current x-ray observations ($L_x \sim 10^{34} - 10^{36}$) erg s$^{-1}$, the duration of activity for magnetars can be maintained for ~ $10^4$-$10^6$ yrs.

We have calculated in detail the x-ray luminosity of magnetars in terms of the electron capture process[10], and making comparison with the observations for magnetars. We note that up to now, there are no other known theoretical models that can fully delineate the observed high x-ray luminosity of the magnetars. However, using the physical mechanism we proposed, the theoretical prediction is basically consistent with the observational data of x-ray luminosity of magnetars[10] (Fig.3).

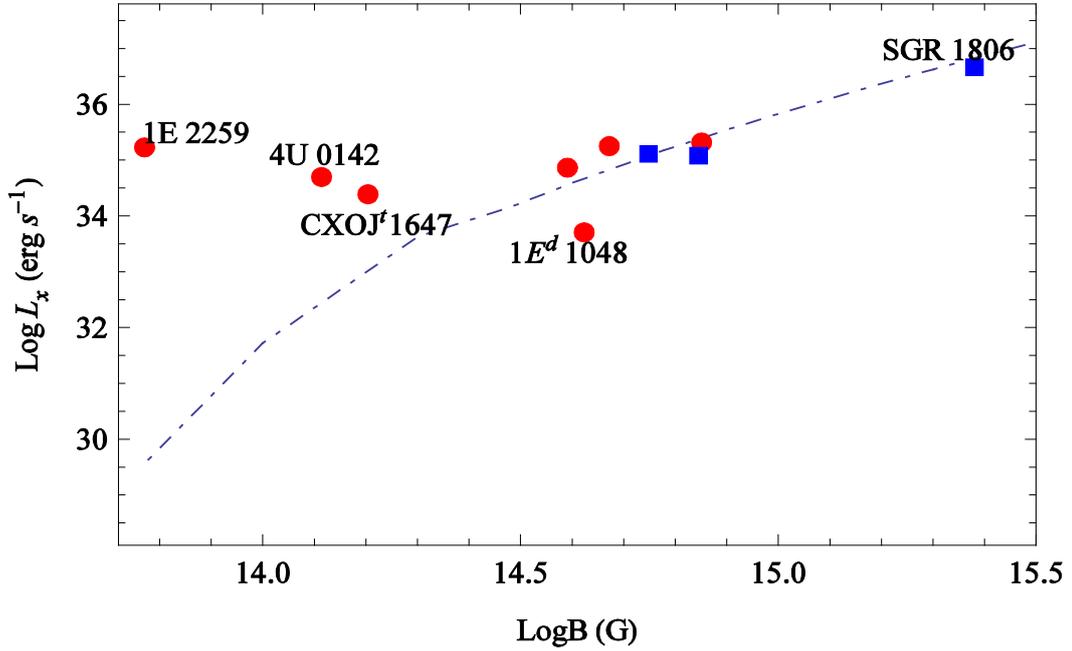

The Soft Gamma Repeaters(SGR) are showed by red cycles; the Anomalous X-ray Pulsars (AXPs)are showed by blue blocks. Some obvious accretion phenomena have been observed for the three AXPs (in the left region) far from the theoretical curve.

**Figure 3.** Comparing the observed magnetar $L_X$ with the calculations.

## 7. Result and Discussion

The important new result for the Fermi energy of a strongly magnetized and relativistic electron gas derived in the last section was applied to explain the high X-ray luminosity of magnetars. It is expected that our new result may have other important applications in the physics of neutron stars and magnetars. For instance, in the absence of magnetic fields it is generally believed that the ordinary URCA process ( $p + e^- \rightarrow n + \nu_e$ ; $n \rightarrow p + e^- + \bar{\nu}_e$ ) can not occur in neutron star interiors unless the proton fraction is greater than 9%. This is because the conservation of both energy and momentum of the leptons in the URCA process can not be simultaneously satisfied under $\beta$-decay equilibrium in neutron star interiors[11]. However, the modified URCA process may occurs ( $n + p + e^- \rightarrow n + n + \nu_e$ ; $n \rightarrow p + e^- + \bar{\nu}_e$ ), but the energy loss rate of the modified URCA process is much smaller than that of the ordinary URCA process (if it could happen). On the other hand, in the presence of the superstrong magnetic fields, both the ordinary( or direct) URCA process and the modified URCA process are possible, and leading to significant neutrino energy loss rates. In particular, in the interiors of nascent pulsars with very strong magnetic fields $10^{12} - 10^{13} Gauss$ in their polar region, there may exist relatively large anisotropic neutron superfluid $^3P_2$ region with ultrastrong magnetic field exceed the critical field strength $B_{cr}$. The Fermi energy of the extreme relativistic electrons(with electron fraction $\approx 5\%$) in these neutron superfluid $^3P_2$ region could exceed the Fermi energy of the nonrelativistic neutrons. This mean that the ordinary( or direct) URCA process can occur in such neutron superfluid region with superstrong magnetic fields. We then anticipate at least two important effects would result and leading to significant astrophysical consequences.

1. It can provide an effective fast cooling mechanism for the young pulsars.
2. The ordinary URCA process happening in the anisotropic $^3P_2$ neutron superfluid region in neutron star interior may provide an effective cooling mechanism. On the other hand in 1982 we proposed a heating mechanism[12] for neutron star interiors based on the magnetic dipole radiation generated by the superfluid vortex motion from the magnetic moment of the $^3P_2$ neutron Cooper pairs. These two mechanism may be combined to generate the anisotropic neutron superfluid with two different phases A and B similar to that of nisotropic liquid helium $^3He$. It is then expected that the oscillation between these two phases of the anisotropic neutron superfluid may be used to naturally explain the observed glitch phenomena of the young pulsars. The research of these topics is now in progress.

## Acknowledgements

We would like to express our sincere appreciation for the financial support from the National Natural Science Foundation of China> No. 11273020.